\documentclass[english,aps,prb,twocolumn,amsmath,amssymb]{revtex4}
\usepackage[T1]{fontenc}
\usepackage[latin1]{inputenc}

\usepackage{graphicx}
\usepackage{epsfig}
\usepackage[usenames]{color}
\usepackage{amsmath}
\usepackage{bm}
\usepackage{ae}
\usepackage{pstricks,pst-grad}
\usepackage{babel}
\makeatother
\makeatletter

\begin{document}

\title{Static and dynamic properties of the interface between\\
       a polymer brush and a melt of identical chains}

\author{C.~Pastorino}
\affiliation{Institut f{\"u}r Physik WA331, Johannes Gutenberg-Universit{\"a}t, 55099, Mainz, Germany}
\author{T.~Kreer}
\affiliation{Institut Charles Sadron, 6 Rue Boussingault, 67083 Strasbourg, France}
\author{M.~M{\"u}ller}
\affiliation{Department of Physics, University of Wisconsin-Madison, WI 53706-1390, USA}
\affiliation{Institut f{\"u}r Theoretische Physik, Friedrich-Hund-Platz 1, 37077 G{\"o}ttingen, Germany}
\author{K.~Binder}
\affiliation{Institut f{\"u}r Physik WA331, Johannes Gutenberg-Universit{\"a}t, 55099, Mainz, Germany}

\begin{abstract}
Molecular dynamics simulations of a short-chain polymer melt between
two brush-covered surfaces under shear have been performed. The end-grafted
polymers which constitute the brush have the same chemical properties
as the free chains in the melt and provide a soft deformable substrate.
Polymer chains are described by a coarse-grained bead-spring model with
Lennard-Jones interactions between the beads and a FENE potential
between nearest neighbors along the backbone of the chains. 
The grafting density of the brush layer offers a way of controlling
the behavior of the surface without altering the molecular interactions.
We perform
equilibrium and non-equilibrium Molecular Dynamics simulations at constant temperature
and volume using the Dissipative Particle Dynamics thermostat. The
equilibrium density profiles and the behavior under shear are studied
as well as the interdigitation of the melt into the brush, the orientation
on different length scales (bond vectors, radius of gyration, and
end-to-end vector) of free and grafted chains, and velocity profiles.
The viscosity and slippage at the interface are calculated as functions
of  grafting density and shear velocity. 
\end{abstract}
\maketitle

\section{Introduction}

Grafted polymer layers have received abiding attention because
coating surfaces with polymer brushes offers an opportunity to control
static and dynamic surface properties. Polymer brush coatings have
been utilized in a wide variety of applications ranging from the stabilization
of colloidal suspensions (e.g., paint), the reversibly tuning of wetting and
adhesion properties to lubrication, friction and wear.\cite{Advincula} For instance,
polymer brushes are renown for their ability to drastically decrease
the friction between two solid surfaces.\cite{jacob_klein2} The work
of adhesion and wettability are also changed when a solid surface
is coated with end-grafted polymer layers. Thus, polymer brushes and
cross-linked rubbers find applications as adhesives.

Polymer brushes are also interesting from a fundamental point of view
because many properties are determined by the conformational entropy
of the extended macromolecules, in particular, the stretching of the
polymers away from the grafting surface \cite{MWC} in response to increasing
the grafting density or swelling the brush with a solvent. Increasing
the grafting density is an experimentally convenient way of altering
the surface properties without altering  molecular interactions.
This gives rise to a rich wetting behavior even in apparently simple
model systems like a polymer drop on top of a brush of identical chains. \cite{Leibler,Matsen,Maas,droplets_marcus_luis2,preprint}
If there are few chains grafted onto the surface, an increase of the
grafting density will result in a smaller contact angle because the
melt chains benefit from the additional attraction provided by the
brush. At intermediate grafting densities this might result in complete
wetting of the melt on the top of brush.
If the grafting density is too high, however, the
polymer melt will dewet from the brush -- a phenomenon termed autophobicity.
Such {}``entropic dewetting'' has been observed experimentally in
polymer networks \cite{kerle} and polymer brushes.\cite{reiter,voronov}
Autophobicity can be traced back to the formation of an interface
between the strongly stretched chains in the polymer brush and the
free chains of the melt. \cite{Leibler,Matsen,Maas,droplets_marcus_luis2}

The interface between brush and melt has many important applications:
Such interfaces commonly occur between two immiscible
homopolymers that have been reinforced by a diblock copolymer. The
diblock adsorbs at the interface as to place its blocks into the respective
phases. Thereby, it lowers the interfacial tension and, for sufficiently
large surface excess, each block forms a polymer brush in contact
with the homopolymer-rich bulk phase. \cite{Leibler_cop,Shull} The interdigitation between
brush and homopolymer chains determines many mechanical properties
of the interface. \cite{Brown,MilnerMRS} 
While the equilibrium properties of these copolymer
brushes at the interface between incompatible homopolymers (as well
as brushes formed from end-adsorbed chains) resemble those of irreversibly
grafted polymer brushes, their dynamic properties might substantially
differ due to the in-plane mobility. Another realization of a brush-melt
interface occurs when a polymer melt dewets from a dense polymer brush
(or rubber \cite{kerle}). These soft deformable substrates might offer
great control over flow properties. As we shall demonstrate below,
changing the grafting properties one can tune the hydrodynamic boundary
conditions. If the grafting density is high enough for autophobic
dewetting to occur,, additional mechanisms of dissipation for moving
droplets like lifting of the three-phase contacts line (wetting ridge) \cite{Shanahan} 
to balance the normal components of the forces on the contact line
or chain pull-out \cite{reiter2} can be envisaged.

The equilibrium properties of polymer brushes in various solvents have
been studied by Monte Carlo \cite{LAI1991,LAI1992} and Molecular
Dynamics (MD) simulations\cite{Grest,Gast,kostas,murat_grest}
as a function of chain length and grafting density. There are fewer
studies investigating the equilibrium properties of the brush-melt
interface. Grest studied the limit in which the melt chains are much
shorter than the chains of the brush.\cite{Grest} Daoulas \emph{et al.} performed
detailed atomistic Monte Carlo simulations to analyze the equilibrium
structure a chemically realistic united atom model for polyethylene
melt in contact with a brush grafted onto a carbon substrate. \cite{kostas}
These brush-melt interfaces of long entangled chains have also been
investigated in the context of adhesion. Sides and Grest used glassy
brush-melt interfaces to study chain pull-out and craze formation
under tensile stress by means of simulations. \cite{sides_grest_adhesion}

Although the molecular mechanisms and the concomitant boundary conditions
(amount of slip) of flow at polymeric interfaces have attracted abiding
interest, it is only incompletely understood. Many distinct physical
processes are involved such as wetting, roughness and bubble formation. \cite{cottine-bizonne_barrat}
An accurate prediction of non-equilibrium properties is important
for the design of nano- and microfluidic devices characterized by
a large surface-to-volume ratio. In particular, the boundary condition
of polymer melts between walls has been extensively studied experimentally, \cite{exp_melt_slip,leger}
by computer simulation \cite{varnik,priezjev} and by analytic theory.
\cite{brochard_degennes,Bocquet} The boundary condition
and friction mechanisms have also been studied experimentally with near
field laser velocimetry techniques, including anchored chains in the
solid surface. \cite{leger} 
The sliding of two brushes and the non-equilibrium behavior of 
brush-brush interfaces can be studied by surface force apparatus (SFA) measurements \cite{jacob_klein2}
and the experimental results have been compared with computer simulations.
\cite{Grest,Gast,Kreer2001,Kreer2003} These simulations emphasize
the correlation between the non-equilibrium shear properties and the
interdigitation and orientation of the two apposed brushes.

In this work we study the behavior of the brush-melt interface of
short non-entangled chains of a coarse-grained model in equilibrium
and under shear as a function of the two experimentally accessible
parameters: grafting density and shear velocity. The grafting density
is systematically varied over a wide range from the mushroom-regime
to high grafting densities that correspond to autophobic behavior.
MD simulations in junction with the Dissipative Particle Dynamics
thermostat are employed to account for hydrodynamic correlations without
including an explicit solvent. 
Unlike  most calculations in the literature we study polymer chains in a poor solvent.
The attraction between monomers allows us to extend this model to studying liquid-vapor interfaces
and the wetting of a melt on top of a brush of identical chains.
The remainder of our manuscript is organized as follows: In 
section \ref{sec:model} we introduce our coarse-grained model and
the simulation technique. In section \ref{sec:Results} the results of
the simulations for the structure, orientations and boundary conditions
of this melt-brush interface in equilibrium and under shear are detailed.
The last sections contains a discussion of our results and an outlook
on future work.

\section{Model and thermostat\label{sec:model}}

We consider a widely utilized, coarse-grained bead-spring model
for polymers. \cite{kremer_grest}
This model has been used for a variety of thermodynamic conditions,
chain lengths and physical regimes such as glasses, melts and dilute
solutions. \cite{review_kroeger} The potential between neighboring
beads along the same polymer is modeled by a finite extensible non-linear
elastic potential (FENE) ,
\begin{equation}
U_{\rm FENE}=\begin{cases}
-\frac{1}{2}k\,\, R_{0}^{2}\ln\left[1-(\frac{r}{R_{0}})^{2}\right] & r < R_{0}\\
\infty & r\ge R_{0}\end{cases},
\end{equation}
where  $R_{0}=1.5\sigma$, the spring constant
is $k=30\epsilon/\sigma^{2}$, and $r=|{\mathbf{r}_{i}}-{\mathbf{r}_{j}}|$
denotes the distance between neighboring monomers. Excluded volume interactions
at short distances and van-der-Waals attractions between segments are described
by a truncated and shifted Lennard-Jones (LJ) potential:

\begin{equation}
U(r)=U_{{\rm LJ}}(r)-U_{{\rm LJ}}(r_{c})\,,
\end{equation}
with
\begin{equation}
U_{\rm LJ}(r)=4\epsilon\left[\left(\frac{\sigma}{r}\right)^{12}-\left(\frac{\sigma}{r}\right)^{6}\right]
\end{equation}
where the LJ parameters, $\epsilon=1$ and $\sigma=1$, define the units of energy
and length, respectively. $U_{{\rm LJ}}(r_{c})$ is the LJ potential evaluated at
the cut-off radius of twice the minimum of the potential:
$r_{c}=2\times2^{\frac{1}{6}}$. Often a completely repulsive LJ
potential, $r_{c}=2^{\frac{1}{6}}$, has been utilized for computational
efficiency. \cite{Kreer2001,Kreer2003,Kreer2004} This choice corresponds to a
polymer brush in a good solvent. Here, we include attractive interactions
between segments and, thus, are able to describe liquid-vapor coexistence,
wetting and droplet formation below the $\Theta$-temperature,
$k_{B}\Theta/\epsilon\sim3.3$. \cite{droplets_marcus_luis1}

The substrates at $z=0$ and $z=D$ are modeled as  idealized flat
and impenetrable walls which interact with the polymer segments via
an integrated Lennard-Jones potential, of the form 

\[
V_{{\rm wall}}(z)=|A|\left(\frac{\sigma}{z}\right)^{9}-A\left(\frac{\sigma}{z}\right)^{3},\]
where $A=3.2\epsilon$ is sufficient to make the liquid wet the
bare substrate.  \cite{droplets_marcus_luis1}

The lateral positions of the grafted chain heads are randomly chosen with a total number of chains given by the desired grafting
density. The distance of the first bead from the grafting surface is $z_1=1.2\sigma$.

In our model wall roughness is completely disregarded. The friction
between the melt of free chains and the surface stems from the grafted
chains only. Our goal is to analyze the experimentally controllable
properties of the substrate provided by the grafted polymer layer
without any additional source of roughness arising from a possible
corrugation of the substrate.

We used a Dissipative Particle Dynamics thermostat (DPD) to simulate at
constant temperature \cite{dpd1,dpd2} and account for hydrodynamic interactions
due to the conservation of total momentum. The equations of motion are formally
the same as used in Langevin dynamics: \cite{understanding,tildesley} 

\begin{eqnarray}
\dot{{\mathbf{r}_{i}}} &=&\frac{\mathbf{p}_{i}}{m_{i}}\\
\dot{{\mathbf{p}_{i}}} &=&{\mathbf{F}_{i}}+{\mathbf{F}_{i}}^{D}+{\mathbf{F}_{i}}^{R}
\label{eq:md_eq_motion}\end{eqnarray}
where ${\mathbf{F}_{i}}^{D}$ and ${\mathbf{F}_{i}}^{R}$ denote the dissipative and
random forces respectively. $m_{i}$ is the mass of the
monomers which is set to unity for all the simulations.  The difference between
DPD and Langevin thermostats is that the random and dissipative forces are
applied in a pair-wise form, where the total forces acting on a pair are equal to
zero and, thus, momentum conservation is obeyed. The expression for the forces are the following:

\begin{eqnarray}
{\mathbf{F}_{i}}^{D}=\sum_{j(\neq i)}{\mathbf{F}_{ij}^{D}} & \,\,;\,\,&{\mathbf{F}_{ij}^{D}}=-\gamma\omega^{D}(r_{ij})(\hat{\mathbf{r}}_{ij}\cdot{\mathbf{v}_{ij}})\hat{\mathbf{r}}_{ij}\\
{\mathbf{F}_{i}}^{R}=\sum_{j(\neq i)}{\mathbf{F}_{ij}^{R}} & \,\,;\,\,&{\mathbf{F}_{ij}^{R}}=\zeta\omega^{R}(r_{ij})\theta_{ij}\hat{\mathbf{r}}_{ij},
\end{eqnarray}
where for each vector ${\mathbf{a}}$ we use a notation
${\mathbf{a}_{ij}}={\mathbf{a}_{i}}-{\mathbf{a}_{j}}$, $\gamma$ is the friction
constant and $\zeta$ denotes the noise strength.  Friction and noise, $\gamma$ and
$\zeta$, obey the relation $\zeta^{2}=2k_{B}T\gamma$ and the concomitant
weight functions satisfy fluctuation-dissipation theorem,
$[\omega^{R}]^{2}=\omega^{D}$. $\theta_{ij}$ is a random variable with zero
mean and second moment $\langle \theta_{ij}(t)\theta_{kl}(t^{'})\rangle
=(\delta_{ij}\delta_{jl}+\delta_{il}\delta_{jk})\delta(t-t^{'})$ and the weight
functions are chosen to be: 
 
\begin{equation}
[\omega^{R}]^{2}=\omega^{D}=\begin{cases}
 (1-r/r_{c})^{2},r<r_{c} & ,\\ 0,\,\,\,\, r\geq r_{c}\end{cases}
\end{equation}
which is a common choice for continuous forces. \cite{dpd3}

The equations of motion (\ref{eq:md_eq_motion}) were integrated using
the velocity Verlet algorithm \cite{tildesley,understanding} with
a time step of $dt=0.0005\tau$ or $dt=0.002\tau$ where $\tau=\sigma \sqrt{m/\epsilon}$
denotes the time unit in Lennard Jones parameters. Using the described
simulation scheme $nVT$, non-equilibrium (applied constant shear
velocity in the walls) and equilibrium simulations were performed.

We studied a melt of 10-bead polymer chains with segment number density close
to $\rho_{\rm melt,coex}\sigma^3=0.61$ between two end-grafted polymer
layers of different grafting densities. A temperature of $k_{B}T/\epsilon=1.68$
was used to reproduce the thermodynamic conditions of previous Monte
Carlo and self-consistent field calculations. \cite{droplets_marcus_luis1,droplets_marcus_luis2}
At this temperature the density, $\rho_{\rm melt,coex}$, of the melt in coexistence 
with its vapor of very low density is not too high and an efficient equilibration of 
the system by Monte Carlo or MD simulations is possible. If we reduced
the temperature further the density of the vapor would decrease further but the
density of the coexisting melt would increase and, thus, significantly slow down 
the simulations.

 \begin{figure}[th]
 \includegraphics[width=5cm]{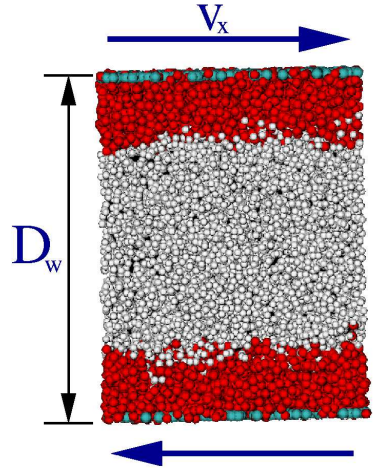}
 \caption{Sketch of the simulation setup. Segments of the free melt chains are
 shown in light grey, the brush is presented in dark grey, while the heads of the brush are
 light. The figure  corresponds to a snapshot of the simulation box for a grafting density of 
 $R_{ee}^2/\Sigma=5.89$ and wall velocity $v_x=1\sigma/\tau$   
 The shear direction, wall velocity and distance between walls are indicated.
 }
 \label{fig:snapshot}
 \end{figure}

The grafting density $\Sigma^{-1}$ denotes the number of grafted chains per
unit area and we measure it in units of the end-to-end distance, $R_{ee}$ of
melt chains in the bulk. $R_{ee}^2/\Sigma \sim {\cal O}(1)$ marks the cross-over
from mushroom to brush regime. The system was simulated for
$R_{ee}^{2}/\Sigma=0.80,\,1.47,\,2.95,\,5.89,\,10.31$, spanning different
physical regimes of the grafted layer -- from the mushroom regime for low
grafting densities to dense, strongly stretched brushes exhibiting autophobic
wetting behavior. Two smaller grafting density values, $R_{ee}^{2}/\Sigma=0.27,\,0.54$,
were also considered to analyze the effective boundary condition of the melt under shear.
The area of the grafting substrate was taken to  be
$21\sigma\times18.2\sigma$, whereas we chose a fixed distance  of
$D_{w}=30\sigma$ between the substrates.  The system was simulated typically
for $1\times10^{6}$ MD steps for thermalization,  while trajectories of length $2.6$ to
$4\times10^{6}$ were used to average data subsequently.  The walls were sheared
relative to each other in the lateral direction, as shown in
Fig.~\ref{fig:snapshot}, where the main parameters of the simulation
are indicated. 

The shear velocity was taken in the range $v_{{\rm x}}=0-2\sigma/\tau$ (or
alternatively shear rates $0 \leq \dot{\gamma}\tau \leq 0.06$). The
equilibrium characteristics of the system were obtained from bulk MD
simulations with periodic boundary conditions in all the spatial dimensions. From them, we extracted the
end-to-end distance, $R_{ee} \equiv \sqrt{\langle R_{ee}^2
\rangle}=3.66\sigma$, and self-diffusion coefficient, $D=0.05\sigma^{2}/\tau$.
The characteristic relaxation time of chain conformations is thus given by $\tau^{*}\equiv
R_{ee}^{2}/D=268\tau$.

To simulate shear we utilized the same thermodynamic conditions as in the equilibrium simulations 
but moved the first beads of the grafted chains with constant velocity, $v_x$,
in the $\hat x$-direction and monitored the force, $F_x$, on the grafting points (see Fig.~\ref{fig:snapshot}).
Using the natural time and length scales, $\tau^{*}$ and $R_{ee}$,
we define a characteristic velocity, $v^{*}=D/R_{ee}=0.014\sigma/\tau$.
The ratio between the shear velocity, $v_x$, measured in units of $v^*$
and the film thickness, $D_w$, measured in units of $R_{ee}$
defines the Weissenberg number, ${\rm We}=(v_x/v^*)(R_{ee}/D_w)=\dot{\gamma} \tau^*$ 
which measures the shear rate in units of the relaxation time of the chains. Our simulations
span a wide range of Weissenberg numbers for this polymer system: $0 \leq {\rm We} \leq 16$.
Adimensional units are used in the remaining sections unless explicitly
mentioned.  Lengths are expressed in units of the chains' end-to-end
distance, $R_{ee}$, in the melt and velocities in terms of the Weissenberg number, We. 
\section{Results\label{sec:Results}}

\subsection{Segment density profiles}

\begin{figure}[th]
\begin{center}\includegraphics[%
  width=7cm]{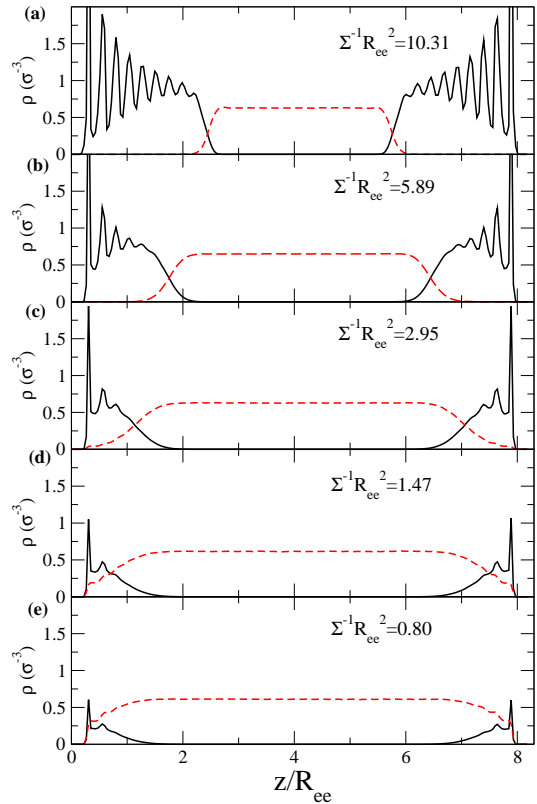}\end{center}
\caption{Equilibrium density profiles for various grafting densities. From
(a) to (e) the grafting densities decrease: $R_{ee}^{2}/\Sigma=10.31,\,5.89,\,2.95,\,1.47,\,0.80$.
Full lines indicate the density of segments belonging to grafted chains
while the segment density of free (melt) chains is marked by dashed
lines. For all simulations,  the melt density is close to the liquid-vapor coexistence density, $\rho_{{\rm melt,coex}}\sigma^3=0.61$
of the melt at the temperature $k_BT/\epsilon=1.68$. \label{cap:density_profiles}}
\end{figure}

First, we illustrate the general structure of the brush-melt interface in
equilibrium and under shear by plotting profiles of the number density of
segments belonging to grafted and free chains in
Fig.~\ref{cap:density_profiles}.  The grafting density decreases from top to
bottom: $R_{ee}^{2}/\Sigma=10.31,\,5.89,\,2.95,\,1.47,\,0.80$. These values
cover the entire regime from strongly stretched brushes to mushroom-like
configurations. Upon increasing the grafting density we observe the building-up of
an interface between the brush and the melt. For low grafting densities,
corresponding to the mushroom regime of the brush, the melt chains reach
 the substrate. For intermediate grafting densities the brush layer starts
to expel the melt chains and their density at the substrate becomes vanishingly
small for $R_{ee}^{2}/\Sigma\geq2.95$. As we increase the grafting density
further, the average density inside the brush exceeds the liquid density,
$\rho_{\rm melt}$, and oscillations in the density indicate fluid-like
layering effects inside the brush. The layering becomes more pronounced when
one increases the brush density and it is somewhat stronger than in
previous simulations\cite{Kreer2001,Kreer2003} because the grafting density
is higher in our case. 

The brush chains stretch and the well-developed interface 
between melt and grafted chains moves farther away from the grafting surface and becomes 
narrower. For the highest grafting density studied, $R_{ee}^2/\Sigma=10.31$, 
there is only very little interdigitation between brush and melt. It is this
interdigitation between brush and melt which is important for the static
and dynamic properties of the system.

\begin{figure}[th]
\begin{center}\includegraphics[%
  width=7cm]{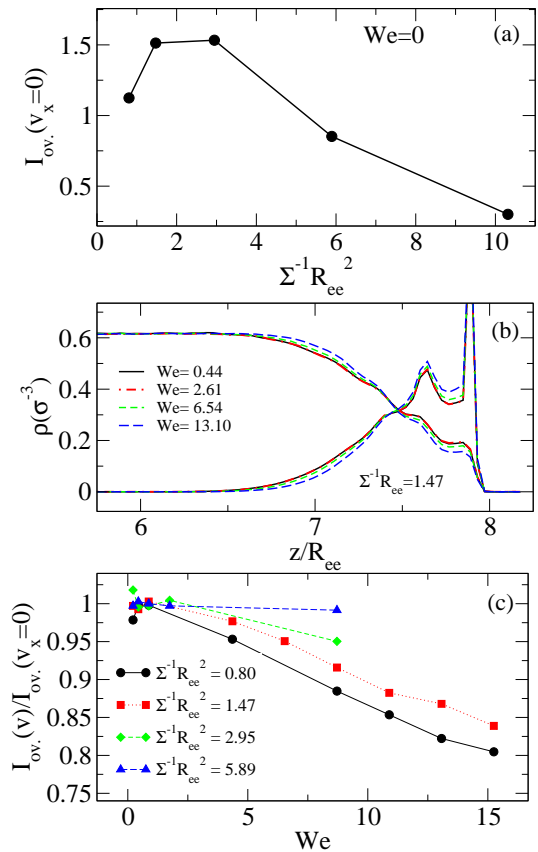}\end{center}
\caption{
Overlap integral  as a function of the grafting density (a).  
Panel (b) presents the response to shear of the density profiles at intermediate grafting
densities. The overlap integral as a funstion of velocity is presented in (c).
\label{cap:Overlap-integral}}
\end{figure}

To characterize the change of the brush-melt interface as a function of
grafting density and shear velocity, we compute the overlap between grafted and
free chains: \cite{Kreer2001}

\begin{equation}
I_{{\rm overlap}}= \frac{R_{ee}^2}{N \rho_{\rm melt,coex}}\int_{0}^{D_w}\rho_{{\rm brush}}(z)\times\rho_{{\rm melt}}(z)dz\,,\label{eq:int_overlap}
\end{equation}
where $\rho_{{\rm brush}}$ and $\rho_{{\rm melt}}$ are the brush and melt
densities, respectively. The integrand in Eq.~(\ref{eq:int_overlap}) is
non-zero only in the regions of the film where there are free and grafted
polymers and the integral quantifies the total amount of interdigitation.
$I_{{\rm overlap}}$ is directly proportional to the number of interactions 
between two polymer layers and therefore provides an insight on how the shear 
stress depends on normal pressure and shear rate. \cite{Kreer2001,Kreer2003}

Fig.~\ref{cap:Overlap-integral}(a) presents the overlap between brush and melt
as a function of the grafting density. As we increase the grafting density, the
overlap between melt and brush increases because there are more grafted
chains at the surface. Upon increasing the grafting density further, however,
we observe that the overlap between melt and brush passes through a maximum and
decreases.  The behavior of the overlap between brush and melt parallels the
wetting behavior, where one observes that the brush favors wetting of the melt
at low grafting densities but leads to autophobic dewetting at large grafting
densities.

Panel (b) shows the response  of the density profiles to shear for an  intermediate grafting
density. In qualitative agreement with simulations of polymer brushes
in a good continuum or monomeric solvent (cf.~Ref.~[\onlinecite{Gast}] and references
therein) the height of the brush decreases upon shear. Here, we observe the
behavior for a brush under bad solvent condition immersed into a polymer melt.
The thinning of the brush goes along with a significant decrease of the
melt-brush interdigitation (cf.~panel (c) of Fig.~\ref{cap:Overlap-integral}),
i.e., a sharpening of the brush-melt interface, and it also correlates with the
inclination of the grafted polymers for low grafting densities (see subsection
\ref{sub:Orientations-and-structure}). For the highest grafting density,
$R_{ee}^{2}/\Sigma=5.89$, however, the density profiles are almost independent
of shear velocity. The chains are so stretched that the shear has less impact on their
conformations.

\subsection{Orientation and structure \label{sub:Orientations-and-structure}}

Next we study conformational and orientation properties. Three distinct
phenomena determine the molecular orientation: The grafting substrate or the
brush melt interface tend to align the extended molecules parallel to the
substrate or interface, respectively. The crowding of the grafted chains at
high grafting density results in chain stretching and a perpendicular
orientation.  Finally, shear tends to align molecules in the direction of the
flow. 

\begin{figure}[th]
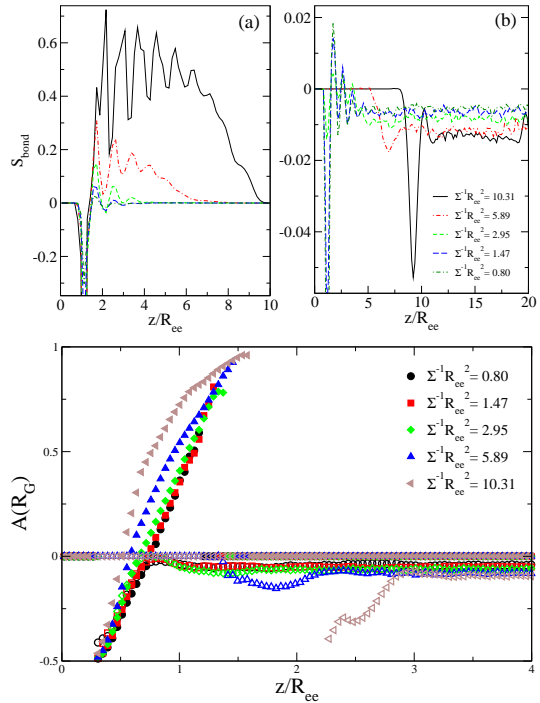

\begin{center}\begin{tabular}{c}
\includegraphics[clip,
  width=7cm]{figures/norm_bond_orient.eps}\tabularnewline
\includegraphics[%
  width=7cm]{figures/Rg_v1.eps}\tabularnewline
\end{tabular}\end{center}
\caption{Orientations of the brush-melt interface on different length scales,
grafting densities and at fixed shear velocity, $v_{x}=1\sigma/\tau$ (or
We=8.71).  Panels (a) and (b) present the bond orientation of the brush and the melt
of free chains, respectively.  Part (c) shows the asphericity parameter
$A(R_{G})$ of the radius of gyration.  Full symbol correspond to the chains in
the brush while open symbols characterize the behavior of melt chains.  
\label{cap:orient_brush_melt}}
\end{figure}

In Fig.~\ref{cap:orient_brush_melt} the orientation of the
grafted layer and melt are shown as a function of the distance, $z$, from
the substrate for various grafting densities. The bond orientation parameter
is defined as 

\begin{equation}
S_{{\rm bond}}=\frac{3\langle \cos^2 \theta \rangle -1}{2},
\end{equation}
where $\theta$ denotes the angle between the vector connecting two bonded segments and the
$\hat{z}$ direction. So defined, $S_{{\rm bond}}\sim1$ characterizes bonds that
mainly orient along the $\hat{z}$ direction and $S_{{\rm bond}}\sim0$
corresponds to randomly oriented bonds. $S_{{\rm bond}}\sim-0.5$
denotes bonds with an average orientation perpendicular to the substrate
(i.e., along the shear direction, $\hat{x}$, if we apply shear). The orientation on
the scale of the entire molecule is described by the 
asphericity order parameter. We define it for the radius of gyration, $\mathbf{R}_{G}$,
and end-to-end vector, $\mathbf{R}_{ee}$, as

\begin{equation}
A(\mathbf{R})=\frac{R_{z}^{2}-\frac{1}{2}(R_{x}^{2}+R_{y}^{2})}{R_{x}^{2}+R_{y}^{2}+R_{z}^{2}},
\end{equation}
where $R_{i}$ (i=x,y,z) denote the different Cartesian components of the considered
vector. 

$S_{{\rm bond}}$ for different grafting densities and for Weissenberg number  
We$=8.71$  is presented in Fig.~\ref{cap:orient_brush_melt}. Panel (a)
presents the behavior of the grafted chains.  In the ultimate vicinity of the
grafting surface bonds align parallel similar to the behavior of a melt in
contact with a hard substrate. The behavior very close to the grafting surface
is largely independent from the grafting density. The bond orientation
parameter exhibits some oscillations that mirror the layering observed in the
segment density profiles and these oscillations become more pronounced at
higher grafting densities (cf.~Fig.~\ref{cap:density_profiles}). At very low
grafting densities (i.e., in the mushroom regime) the bond orientation
parameter rapidly decays for larger distances from the surface. At intermediate
and large grafting densities, however, there is a pronounced perpendicular
orientation of the bond vectors in the middle of the brush which stems from the
stretching of the grafted chains. At the top of the brush the orientation
parameter decays to zero.

Part (b) of Fig.~\ref{cap:orient_brush_melt} shows the corresponding
orientation of free chains in the melt.  In the ultimate vicinity of the
grafting surface we find a pronounced parallel orientation similar to the first
segments of the brush. Otherwise, the magnitude of the orientation of bond
vectors in the melt is much smaller than for brush chains. There is only a
small parallel alignment of the bond vectors that arises from the shear. It is
slightly less pronounced inside the brush than in the middle of the film and
there is a slight increase of alignment as we increase the grafting density.
This can be rationalized as follows: The larger the grafting density, the
higher is the velocity gradient in the middle of the film and the concomitant
orientation (cf.~the discussion of the slip length in subsection \ref{slip}).
%Most notably, the bond orientation parameter of free chains shows a minimum
%(i.e., parallel orientation) which becomes more pronounced as we increase the
%grafting density and whose location correlates with the overlap region between
%brush and melt. 
The behavior is similar to the parallel alignment at a
polymer-polymer interface or a substrate. It signals the formation of a
well-defined brush-melt interface at high grafting densities and the difficulty
of the free chains to penetrate into the brush which leads to autophobicity.

In panel (c) we investigate the behavior on larger length scales. Not
surprisingly, orientation effects are much more developed for the radius of
gyration than for bond vectors but the qualitative behavior is similar.
  Grafted chains with their centers of mass close to
the grafting surface adopt pancake-like conformations that are more extended
parallel to the wall than in the perpendicular direction.  The majority of grafted
chains have their mass centers in the middle of the brush region around $z
\approx R_{ee}$ and they are extended away from the grafting surface. The
effect increases for higher grafting densities and the further the chain's
center of mass is displaced from the surface. Free chains in turn, are much
less aligned and their parallel alignment arises from the shear, except for the two
highest grafting densities, where one additionally observes a parallel alignment
of chains at the brush-melt interface.

\begin{figure}
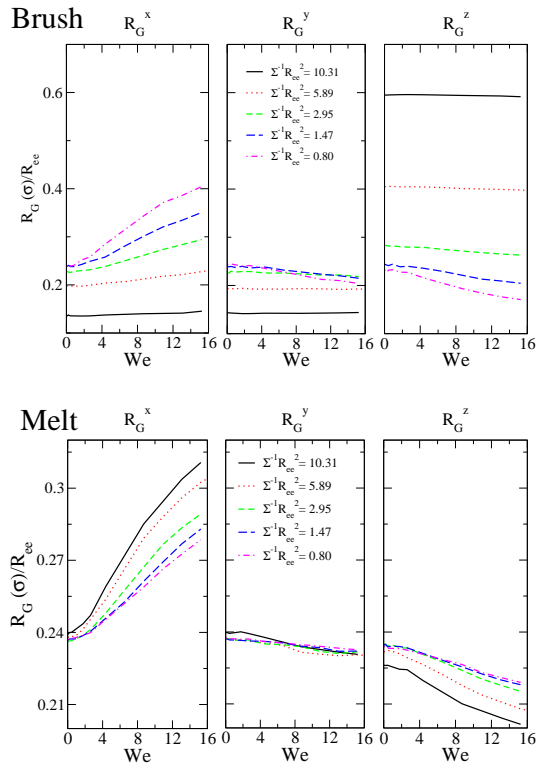

\includegraphics[%
  width=7cm]{figures/mean_sqrt_Rg2_brush_all_rho.eps}\\[5mm]
\includegraphics[%
  width=7cm]{figures/Rg_xyz_all_rhos.eps}
\caption{Cartesian components of the average radius of gyration as function of shear velocity, $v_x$, 
for grafted chains (top) and free chains of the melt (bottom). Different
lines correspond to various grafting densities as indicated in the key.
For a Gaussian chain each Cartesian component of the radius of gyration 
is given by $R_G^i=\frac{R_{ee}}{3\sqrt{2}}$.
\label{fig4}}
\end{figure}

In Fig.~\ref{fig4} we corroborate our findings by analyzing the behavior of the
Cartesian components of the radius of gyration as a function of shear velocity.
The data are averaged over all chains in the film independently of their
distance from the surface.  The top panel (a) shows the behavior of the brush.
As we increase the grafting density the chains stretch away from the grafting
surface. At high grafting density the chain conformation in the brush hardly
depends on shear velocity. At lower grafting density, shear reduces the
stretching perpendicular to the substrate but elongates the chains in the shear
direction. The chain extension in the neutral direction, $\hat y$, is largely
unaffected by shear but decreases as a function of grafting density. This effect
is particularly pronounced for the rather short chains, $N=10$, utilized in our
simulations because the chain extension in the strongly stretched state is not
very much smaller than the contour length and, thus, stretching reduces the
fluctuations of the chain extension perpendicular to the stretching direction.
For longer chains, however,  we would expect
a decoupling between the chain extension parallel and perpendicular to the
substrate and, consequently, a smaller influence of the grafting density on
$R_{G}^y$.

The behavior of the free chains is presented in panel (b). The dependence of
the chain dimensions on the grafting density is rather weak. At vanishing shear
velocity and high grafting density the chains are slightly aligned parallel to
the brush-melt interface ($x$-$y$-plane). Fig.~\ref{cap:orient_brush_melt} suggest
that  similar orientation effects would also arise at the substrate for low grafting 
densities but there are
very few melt chains in the film whose center of mass is in the vicinity of the
grafting surface. Upon increasing the shear rate, we observe an orientation of
the free chains in the shear direction.

\begin{figure}
\begin{center}\includegraphics[%
  width=7cm]{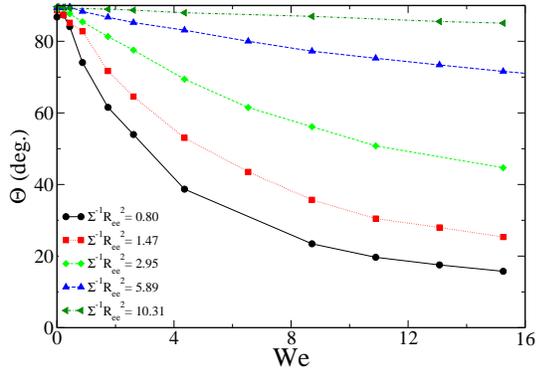}\end{center}
\caption{Average angle of the radius of gyration for the brush layer as a function
of wall velocity for different grafting densities. $\Theta=90$ corresponds to a brush completely
perpendicular to the wall. \label{fig:angle}}
\end{figure}

It is of interest to study to what extend the conformational changes of brush
and melt chains arise from orientation of the chains or from deformation of the
chains. In Fig.~\ref{fig:angle} the average angle of the end-to-end vector of
grafted chains with respect to the grafting surface is shown. 
This is defined as:
\[
\Theta = 90^o - \arccos(\vec{R}_{ee} \cdot\hat{z}/R_{ee}),
\]
where $\vec{R}_{ee}$ is the mean end-to-end vector and $\hat{z}$ is the direction perpendicular
to the wall.
Without shear,
the average $x$ and $y$ coordinates of the brush segments coincide with the
location of the grafting point in the $x$-$y$-plane and this configuration
corresponds to an angle of $90^0$. Upon shearing the film the chains tilt and
the average angle decreases. The angle is smaller, the larger the shear rate
and the lower the grafting density are chosen. For small shear rates the angle
decreases roughly linear with the shear velocity, $v_x$, and saturation effects
become apparent once the angle drops below $45^0$.

\begin{figure}
\begin{center}
\includegraphics[clip, width=7cm]{figures/orient_melt_center.eps}\\
\includegraphics[clip, width=6.5cm]{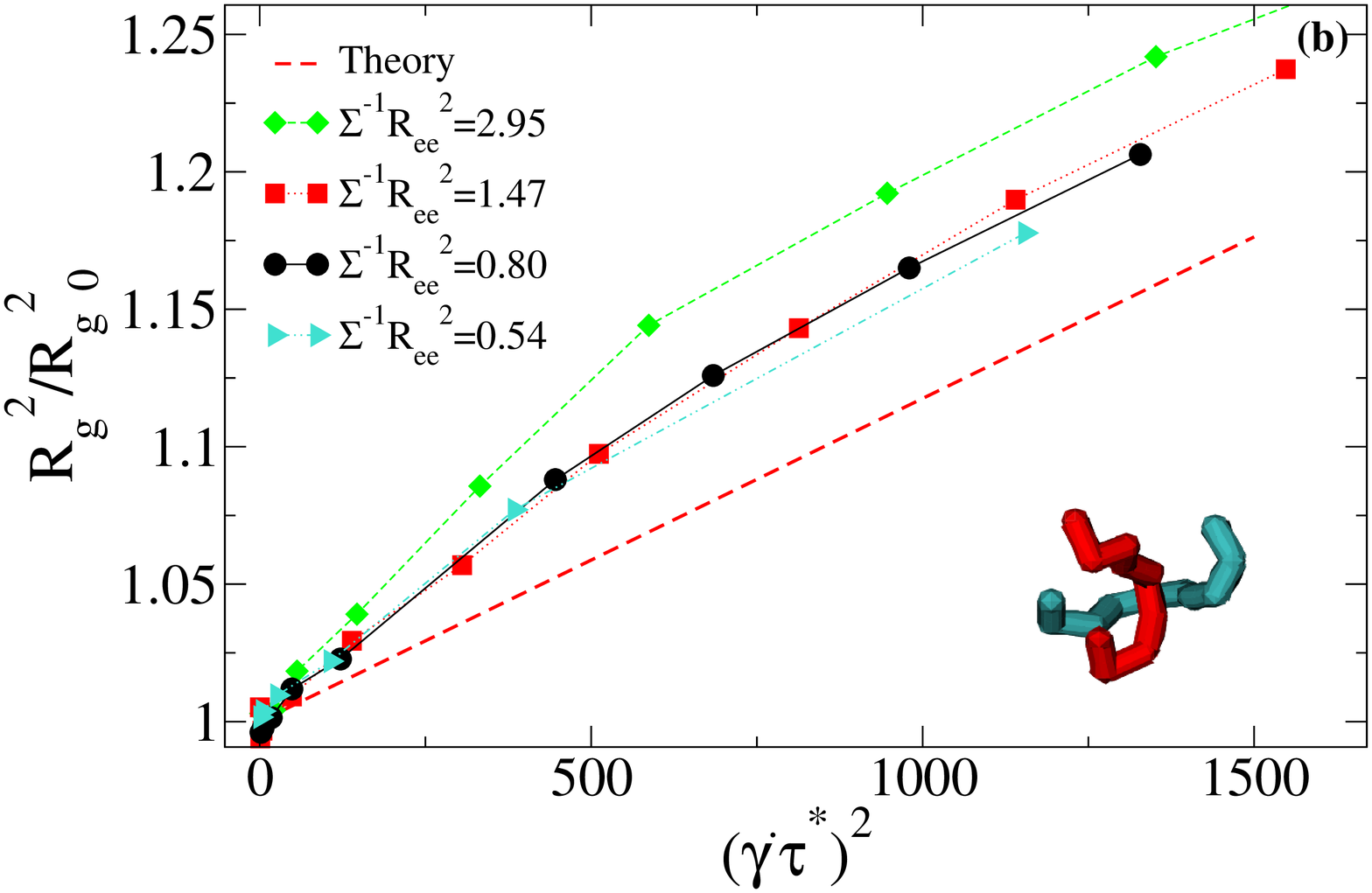}
\end{center}
\caption{(a) Orientation of the free chains in the center of the film versus
Weissenberg number.  The different panels present the bond
orientation parameter $S_{\rm bond}$ (top), the asphericity parameter for
radius of gyration $\mathbf{R}_{G}$ (center), and end-to-end vector
$\mathbf{R}_{ee}$ (bottom) for various grafting densities as indicated in the
key, respectively.\label{fig:Amiddle}
(b) Deformation of free chains in the center as a function of the adimensional shear rate. The
theoretical results by Bruns\cite{Bruns} is indicated with dashed line. An example of the topological
arrangement of the chains is shown in the inset.
}
\end{figure}

 In Fig.~\ref{fig:Amiddle} (a), the orientation of the free chains in the middle of the film is shown for
different length scales. The alignment with the shear
direction is the more pronounced the larger the length scale and, analog to the
behavior of the brush, the orientation effects depend roughly linearly on the
shear rate. 

This orientation effect contrasts with the deformation of the chain which can
be measured by the change of the mean square radius of gyration.  In
Fig.~\ref{fig:Amiddle} (b) we find that the coil deformation depends quadratically on
the shear rate  for small to intermediate shear velocities and a saturation appears for the highest shear rates.  The data are averaged over the central region of the film
considering  a width of $1.4 R_{ee}$. The shear rate is also evaluated there to assure a linear velocity profile, 
for a precise calculation of the shear rate. As showed 
in Fig.~\ref{fig:Amiddle}, the curves for different grafting densities collapse,  except for the $\Sigma^{-1}R_{ee}^{2}=2.95$. This is expected due to the fact that, for this case, the melt density 
is not exactly $\rho\sigma^{3}=0.61$, as mentioned previously.  The quadratic behavior finding is in accord with the theoretical 
prediction\cite{Bruns}:
\begin{equation} 
R_G^2(\dot{\gamma}\tau)=R_G^2(0) (1+ \frac{(\dot{\gamma}\tau)^2}{8508}),
\end{equation} 
where $R_G(0)$ is the radius of gyration at 0 shear rate. However, we see a  difference with the  slope predicted theoretically  in the context of the Rouse model\cite{doi_edwards}. We  attribute this to the topological interactions
of the chains even though they are not long enough to be entangled (see an example taken from the simulation in 
the inset of ~\ref{fig:Amiddle}(b)). 

%% whose result in the context of Rouse model\cite{doi_edwards} is also shown in   Fig.~\ref{fig:Amiddle}. The difference in the slope

\subsection{Viscosity of the melt}
\begin{figure}[th]
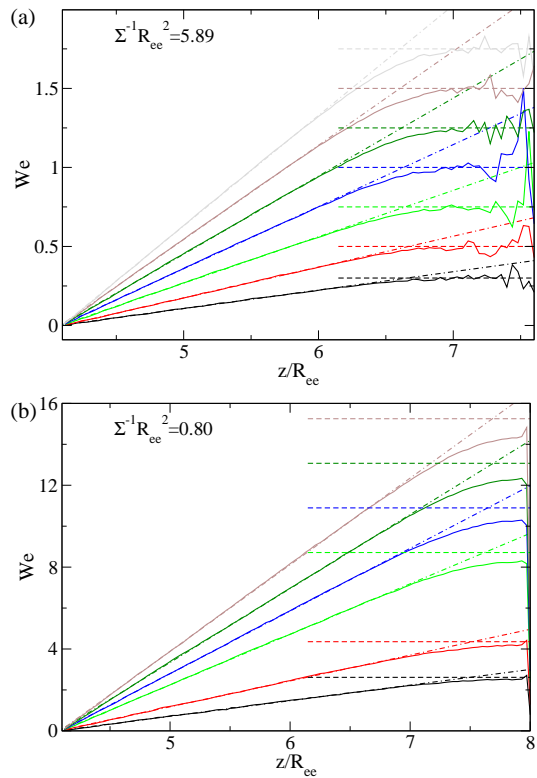

\begin{center}\begin{tabular}{c}
\includegraphics[%
  clip,
  width=7cm]{figures/vel_prof_melt_half_0.44_final.eps}\tabularnewline
\includegraphics[%
  clip,
  width=7cm]{figures/vel_prof_melt_half_0.06_final.eps}\tabularnewline
\end{tabular}\end{center}
\caption{Velocity profiles as function of wall velocity for: (a) grafting
density $R_{ee}^{2}/\Sigma=5.89$ and (b) $R_{ee}^{2}/\Sigma=0.80$.
The wall (and brush layer) velocity is indicated with dashed horizontal lines.
Dashed-dotted lines show the linear extrapolation of the far-field velocity.
In the first case stick boundary condition is observed for the fluid
in the melt-brush interface. The latter case corresponds to a finite
slip boundary condition in which the fluid never reaches the brush
layer velocity. \label{cap:velocity_profile}}
\end{figure}

In Fig.~\ref{cap:velocity_profile} the velocity profiles of the melt under
steady shear are presented.  From the velocity profile of the melt in the
center of the film we extract the shear viscosity, $\eta$, of the melt bulk phase according 
to
\begin{equation}
\eta=\frac{\left\langle F_{x}\right\rangle /A}{|\nabla v_x|},
\end{equation}
where $\left\langle F_{x}\right\rangle$ denotes the mean force of the walls in
the shear direction and $A$ is the area of the substrate. $\nabla v$ is the 
velocity gradient at the center. We use the linear velocity profile in the
middle of the film (cf.~Fig.~\ref{cap:velocity_profile}) to extract the
gradient according to $|\nabla v_x| = \frac{v_x}{D_w/2+b}$, where $D_w$ indicates 
the film thickness and $b$ the slip length (see subsection \ref{slip} ).

\begin{figure}[t]
\begin{center}
%not anymore \includegraphics[%
%not anymore   width=9cm]{figures/eff_visc_v_melt.eps}\tabularnewline
\includegraphics[%
  clip,
  width=9cm]{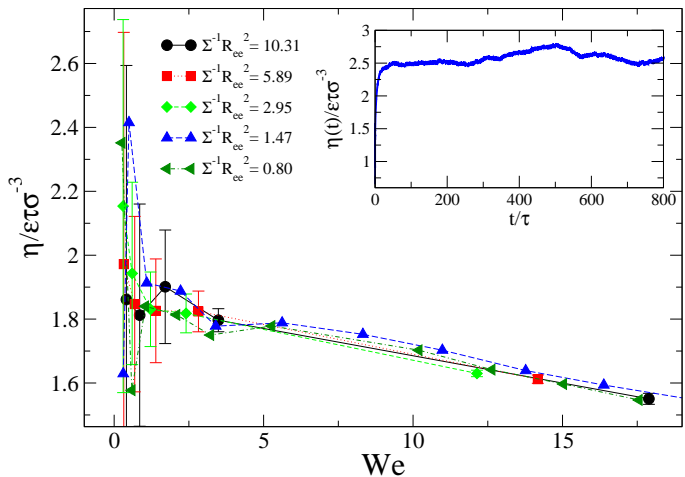}\tabularnewline
\end{center}
\caption{ Viscosity of the melt as a function of the shear velocity for
the studied grafting densities. The melt density, indicated in the
legends was set to $\rho=0.61\sigma^{-3}$.
The inset shows the viscosity as a function of $t$ as given by the
Green Kubo formula. \label{cap:Viscosity}}
\end{figure}

The simulations were performed at constant volume and therefore the final value
of the melt density in the middle of the film is unknown ahead of time and
depends on how much the brush layer squeezes the melt. Consequently, we
performed a first set of runs where the density in the melt slightly
deviated from the target value, $\rho \sigma^3=0.61$, and then adjusted the
number of chains in the simulation cell. Alternatively, we could have simulated
at constant normal pressure. The results of the simulations with constant density 
in the middle of the film are shown in figure  Fig.~\ref{cap:Viscosity} for the
grafting densities $R_{ee}^{2}/\Sigma=2.95,\,5.89$ and $10.31$.
We observe {\em shear thinning} typical of non-Newtonian fluids in the
studied range of velocities which becomes very pronounced at even higher shear
rates. In the regime of very low shear rates, the data have rather large
error bars because of the very high fluctuations of force in the shear direction.
The error bars shown in the figure are
obtained from a block average analysis\cite{understanding} of the wall force,
$F_{x}$.  This low signal-to-noise ratio makes it difficult an extrapolation to the
viscosity at vanishing shear rate.

A more reliable estimate of the viscosity at zero shear rate can be obtained
from the Green-Kubo formula.\cite{hansen-mcdonald} To this end we performed
simulations of the bulk system with periodic boundary conditions in three
dimensions at density $\rho_{{\rm melt,coex}}\sigma^3=0.61$. The viscosity can be
calculated from equilibrium simulations as the time integral of the pressure
tensor correlation function: \cite{hansen-mcdonald}

\begin{equation}
\eta_{\rm GK}(t)=\frac{V}{k_{B}T}\int_{0}^{t} {\rm d}t'\; \left\langle P^{\alpha\beta}(t') \cdot P^{\alpha\beta}(0)\right\rangle ,
\label{eq:visco_GK}\end{equation}
where $V$ is the volume of the system and the pressure tensor $P^{\alpha\beta}$ is 
given by:

\begin{equation}
P^{\alpha\beta}= \frac{1}{V} \left(\sum_{i}mv_{i}^{\alpha}v_{i}^{\beta}+\sum_{i<j}F_{ij}^{\alpha}r_{ij}^{\beta} \right)\;\;.
\end{equation}
Since the pressure tensor is a collective quantity 10 independent simulation
runs with $4 \times 10^6$ steps each ($dt=0.002\tau$) were needed for an accurate
result.  The average, $\eta_{\rm GK}(t)$ in Eq.~(\ref{eq:visco_GK}),  over the 10
independent runs is displayed in the inset of Fig.~\ref{cap:Viscosity}.  Our
final estimate for the bulk viscosity is $\eta_{\rm GK}=2.5(3)$.  Within the
error bars this value is compatible with the extrapolation of the
zero-shear-rate viscosity from the non-equilibrium simulation. 
It seems  that
the estimates of the non-equilibrium simulations are systematically lower than
the results from the bulk fluctuation analysis. If there really is a systematic
discrepancy, it might be attributed to the 
rather small film thickness in the shear simulations. Especially at high 
grafting densities, where the deviations are the largest, the ``bulk-like''
middle of the film is rather narrow and the brush-melt interface tends to align
the chains in the direction of the shear resulting in a somewhat lower estimate
for $\eta$.

\subsection{Effective boundary condition and slip length \label{slip}}
In order to exploit polymer brushes as soft, deformable surface coatings in
micro- and nanofluidic devices two key characteristics of the brush have to be
controlled: (i) the wetting properties of the liquid on top of the brush and
(ii) the boundary condition which describes the velocity field of the fluid in
the vicinity of the surface. Typically, one encodes the microscopic structure
and dynamics at the surface in a single parameter, the slip length $b$, which
is defined as the distance, $z=-b$, behind the substrate where the extrapolation of
the linear, far-field velocity profile, $v(z)$, attains the substrate velocity,
$v_x$, as assumed by a macroscopic no-slip boundary condition. A positive value
of the slip length, $b$, implies that the fluid velocity at the solid surface
does not reach the substrate velocity, i.e., there is slip at the grafting surface,
and the extrapolated velocity profile reaches the surface velocity, $v_x$, inside 
the solid substrate.

The velocity profiles in Fig.~\ref{cap:velocity_profile} also provide
information about the hydrodynamic boundary condition.  The velocity of the
brush segments coincides with the velocity of the wall, $v_x$, of course, and
it is indicated by horizontal lines. For the free chains of the melt we observe
two different behaviors as a function of grafting density:

For medium to higher grafting densities ($R_{ee}^{2}/\Sigma=2.95,\,5.89,\,10.31$) the
liquid flow on top of the brush is described by a {\em stick} boundary condition, 
in which the melt reaches the wall (brush) velocity for all shear rates studied.\cite{comment1}   
For instance, in Fig.~\ref{cap:velocity_profile}(a) the velocity profile of a liquid over a
densely grafted brush, $R_{ee}^{2}/\Sigma=5.89$, is shown for different shear
rates. As one can observe, the boundary velocity is, in fact, attained before 
the substrate and free chains that are inside the brush are completely entrained.
This corresponds to a negative slip length and characterizes the hydrodynamic 
thickness of the brush. \cite{Gast}

\begin{figure}[th]
\begin{center}\includegraphics[clip,
  width=7cm]{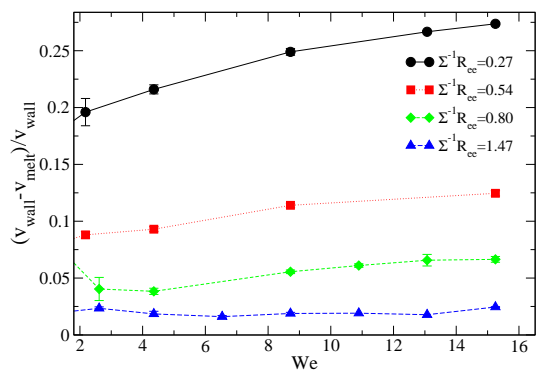}\end{center}
\caption{Relative slip velocity of the melt as a function of shear velocity
for those grafting densities that have partial slip boundary conditions.
\label{cap:slip_vs_v}}
\end{figure}

At lower grafting densities ($R_{ee}^{2}/\Sigma=0.27,\,0.54,\,0.80,\,1.47$), however, the
boundary condition of the melt changes toward a {\em finite slip.} The melt
never reaches the velocity of the grafting substrate, and the slip at the
substrate increases as we decrease the grafting density.  In
Fig.~\ref{cap:slip_vs_v} the slip velocity of the melt at the substrate is
shown as a function of shear velocity for the smallest grafting densities.
The boundary condition for these grafting densities that results in partial slip was
also studied as a function of time. Often experiments report stick-slip
boundary condition  \cite{leger}, where the fluid motion switches from complete
stick to complete slip in the course of time.  We did not observe this
phenomenon in the time window considered in the simulations.  The system rather
exhibits a steady partial slip as observed from the time dependence of the shear
force. 

Figure \ref{cap:velocity_profile}(b) displays the velocity profile for various
shear velocities.  Although there is a notable microscopic slip at the
substrate the concomitant slip length, $b$, extrapolated from the far-field
velocity profile is very small and negative because the velocity profile strongly deviates
from the linear behavior assumed in the extrapolation of the far-field
velocity. Thus, microscopic slip velocity and distortion of the velocity
profile at the substrate compensate each other and the {\em effective} boundary
condition corresponds to no slip.

In the limit of extremely low grafting densities ($R_{ee}^{2}/\Sigma=0.27$), we
observed a positive slip length (see inset of Fig.~\ref{fig:slip_length}): Although the liquid wets the solid
substrate, in the limit of no grafted chains, $R_{ee}^2/\Sigma=0$,
there is complete slip at the substrate (i.e., $b \to \infty$) because it is
perfectly flat.

\begin{figure}[th]
\begin{center}
%\includegraphics[clip,
%  width=7cm]{figures/vel_1d_paper.eps}\end{center}
\begin{tabular}{c}
\includegraphics[clip, width=6.5cm]{figures/vel_1d_paper.eps}  \tabularnewline
\includegraphics[clip, width=7cm]{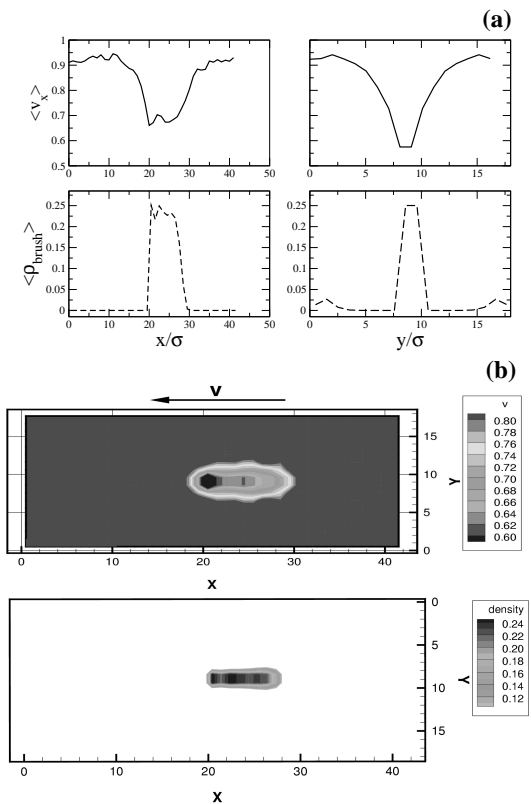} \tabularnewline
\end{tabular}
  \end{center}
\caption{
 (a) Average  melt velocity in the shear direction (plain line) and  density of beads belonging to a grafted chain (dashed line). The left panel shows the profile along the shear direction and the right one in the perpendicular one.  
This simulation was performed with one chain grafted in the surface. Panel (b) shows the two dimensional profile of melt velocity (up) and bead density (down) in a plane 
close to the grafting position $z\sim1.2\sigma$. The center correspond to the position in which the chain is grafted. The shape and extension
of the influence of the grafted chain in the melt velocities can be clearly observed. 
\label{fig:single_brush}}
\end{figure}

\begin{figure}[th]
\begin{center}\includegraphics[clip,
  width=7cm]{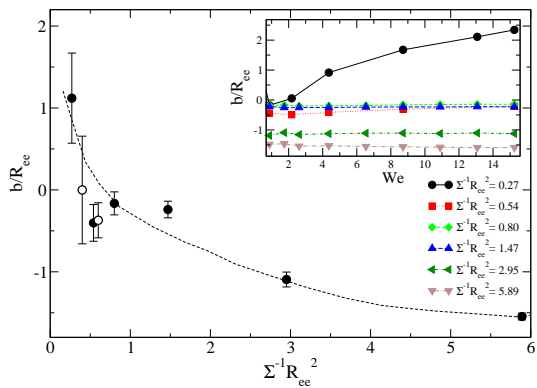}\end{center}
\caption{Slip length as a function
of grafting density for various grafting densities, that 
give rise to  stick and partial slip boundary conditions. The points indicated with open circles 
correspond to runs made only for We$=8.71$ and the dashed line is a guide for the eyes. 
The slip length $b$  versus Weissenberg number is presented in the inset.  \label{fig:slip_length}}
\end{figure}

We performed   simulations of a single chain grafted to the wall to analyze the  
influence and spacial extension of the perturbation in the melt fluid velocity. The  capacity
of a single chain to drag is isolated in this way. The comparisson of the mean fluid velocity 
in the shear direction  is presented in Fig.~\ref{fig:single_brush} for a wall velocity of We=8.71. The system
of coordinates was taken to be in the wall, therefore a smaller velocity means that the grafted chain is dragging the
fluid along. Far from the grafted chain a high velocity is observed. 
Panel (a) shows the density of beads belonging to  the grafting chain and the decay of the fluid velocity along and also perpendicular to the shear direction. The asymmetry, with enhancement in the shear direction is due
to the deformation of the grafted chain  in this direction, as can be clearly observed in Fig.~\ref{fig:single_brush}(b) for
a plane close to the grafting point.


The mean slip length, $b$,  as a function of grafting density and Weissenberg
number is plotted in Fig.~\ref{fig:slip_length}. $b$ is rather insensitive to the shear rate, except for
the lowest grafting density considered, $R_{ee}^{2}/\Sigma=0.27$, that shows a clear dependence with the shear rate.
At very high grafting densities,
the slip length is negative and its absolute magnitude is comparable but 
slightly smaller than the thickness of the dense brush (cf.~Fig.~\ref{cap:density_profiles}). At high grafting 
densities the slip length does not strongly depend on the shear rate in 
accord with the insensitivity of the density profiles in this regime.

The errors in the slip length in Fig.~\ref{fig:slip_length} increase significantly 
for small grafting densities. Note that the chains are grafted randomly
and for low grafting densities (mushroom regime), when the grafted chains do not strongly overlap, the influence
of the grafted chains in the fluid is not homogeneous and quenched fluctuations of  the grafting points become important
(a limiting case is showed in  Fig.~\ref{fig:single_brush}
for a single grafted chain). This gives rise to sample-to-sample fluctuations which can be observed via 
the slight asymmetry of the velocity profile across the simulation cell. 
Therefore the error is large at low grafting densities at low grafting densities.

As we reduce the grafting density the slip length increases. In this regime,
$b$ characterizes the rather intricate interplay between different factors: (i)
As the grafting density decreases the physical thickness of the brush decreases
leading to less negative values of $b$. (ii) Additionally, the melt chains
penetrate deeper into the brush (cf.~Fig.~\ref{cap:Overlap-integral}) resulting
in a hydrodynamic thickness that is smaller than the physical thickness. \cite{Gast} (iii) At low
enough grafting densities there is microscopic slip corresponding to an
incomplete entrainment of the melt by the brush.  Eventually, the slip length
will diverge as the grafting density vanishes because of our choice of a
perfectly smooth substrate that exhibits complete slip. 

These results suggest  that the boundary condition of the liquid flow on top of
the brush can be controlled by changing only the grafting density, which is readily
accessible in experiments.  This offers opportunities for studying flows whose
boundary conditions are purely entropically tunable without any change in the
chemical nature of the macromolecules or the substrate.

\section{Conclusions\label{sec:Conclusions}}

In this work, the interface between a brush and a melt was investigated in
equilibrium and under shear as a function of the sliding velocities of the
walls and for a wide range of grafting densities of the brush layer. The
polymers of both, brush and melt, were taken to be physically identical and
described by a coarse-grained bead-spring model. Polymer segments attracted
each other corresponding to a bad solvent below the $\Theta$-temperature. We
used the DPD thermostat to duly account for the hydrodynamic interactions in
the system. 

The building-up of the brush-melt interface was observed and characterized upon
increasing grafting densities: As expected, free melt chains were progressively
expelled from the polymer brush layer as we increase the grafting density. The
melt chains orient parallel to the brush-melt interface and bond vectors align
much less than the end-to-end vector.  This behavior is similar to what is
observed at interfaces between immiscible polymers but here it occurs at an
interface between chemically identical species and is solely due to entropic
effects. The pronounced structural changes at the brush-melt interface also
suggest that there is a free energy cost associated with this interface which
will lead to autophobicity, i.e. the dewetting of the melt of identical chains
from the brush. This was confirmed by setting up a droplet on the top of the
brush layer which proved to be stable  exactly for the same thermodynamic
parameters than those used in the shear simulations. The behavior of droplets
on this autophobic substrate will be the subject of a future publication.

We complement the equilibrium information with the response of the brush-melt
interface to shear. From the gradient of the velocity at the center of the
film and  via correlations of the pressure tensor in equilibrium we have
estimated the shear viscosity of the bulk phase. The density profiles also provide information
about the hydrodynamic boundary condition.
For low grafting
densities a partial-slip boundary condition was found and the interface
structure is strongly affected by shear. At larger grafting densities, however,
the simulation data are well describable by a stick boundary condition and the
brush is only very little affected by shear \cite{comment1}. Stick-slip phenomena could not be
detected during the time window of our simulation. 
 Thus, a dense brush is
efficient in dragging along the polymer melt even though thermodynamically it
would dewet (at coexistence pressure). 
This is a first step towards describing the
motion of droplets on this soft, deformable brush substrate. 
One key result of our study is the ability to change the
thermodynamic properties and hydrodynamic boundary conditions of a melt in
contact with a substrate by changing only the grafting density which is an
experimentally accessible parameter.  We note that the cross-over towards
autophobic behavior and from slip to stick boundary conditions occurs when the
chains of the brush begin to overlap, i.e. at rather moderate grafting
densities. 

\subsection*{Acknowledgments}

It is pleasure to thank Luis Gonzalez MacDowell, J{\"o}rg Baschnagel and
Joachim Wittmer for useful discussions. Financial support by the DFG
within the priority program {}``Micro- und Nanofluidik'' Mu 1674/3-1
and the ESF-program STIPOMAT are gratefully acknowledged. Computing
time was provided by the NIC, J{\"u}lich, Germany.


\begin{thebibliography}{99}
\bibitem{comment1} At extremely high grafting densities we expect the interdigitation between brush and melt to
become so small (c.f.  Fig.~3a) that the finite slip might occur.
\end{thebibliography}
\end{document}